\documentclass[11pt]{article}
\usepackage[T1]{fontenc}
\usepackage{lmodern}
\usepackage[margin=1.05in]{geometry}
\usepackage{amsmath,amssymb,amsthm}
\usepackage{algorithm,algpseudocode}
\algrenewcommand{\algorithmicrequire}{\textbf{Input}}
\algrenewcommand{\algorithmicensure}{\textbf{Output}}
\algrenewcommand{\alglinenumber}[1]{\scriptsize #1}
\usepackage{microtype}
\usepackage{xcolor}
\usepackage[colorlinks=true,linkcolor=blue!50!black,citecolor=blue!50!black,urlcolor=blue!50!black]{hyperref}
\hypersetup{pdftitle={A Polynomial Algorithm for Stochastic Mean-Payoff Games with a Fixed Number of Random Positions},pdfauthor={Tianhang Lu, Han Xiao, Qizhi Fang}}
\newtheorem{theorem}{Theorem}
\newtheorem{lemma}{Lemma}
\newtheorem{proposition}{Proposition}
\newtheorem{corollary}{Corollary}
\theoremstyle{definition}

\theoremstyle{remark}

\newcommand{\R}{\mathbb R}
\newcommand{\Q}{\mathbb Q}
\newcommand{\Z}{\mathbb Z}
\newcommand{\E}{\mathbb E}
\newcommand{\poly}{\operatorname{poly}}
\newcommand{\sgn}{\operatorname{sgn}}
\newcommand{\one}{\mathbf 1}
\newcommand{\norm}[1]{\left\lVert#1\right\rVert}
\newcommand{\abs}[1]{\left\lvert#1\right\rvert}
\newcommand{\clip}{\operatorname{clip}}
\newcommand{\Slots}{\mathcal S}
\newcommand{\EE}{\mathbb E}

\makeatletter
\newcommand{\institute}[1]{\gdef\@institute{#1}}
\newcommand{\email}[1]{\texttt{#1}}
\renewcommand{\maketitle}{%
  \begingroup
    \renewcommand{\thefootnote}{\fnsymbol{footnote}}%
    \renewcommand{\and}{\qquad}%
    \begin{center}
      {\LARGE\@title\par}
      \vspace{1em}
      {\large\@author\par}
      \vspace{0.6em}
      {\small\@institute\par}
    \end{center}
    \@thanks
  \endgroup
  \setcounter{footnote}{0}%
  \global\let\@thanks\@empty
  \vspace{0.6em}}
\makeatother

\title{A Polynomial Algorithm for Stochastic Mean-Payoff Games \\ with a Fixed Number of Random Positions\thanks{Supported in part by the National Natural Science Foundation of China (No.\,12671377) and Shandong Provincial Natural Science Foundation (No.\,ZR2025MS104).}}
\author{Tianhang Lu \and Han Xiao \and Qizhi Fang}
\institute{School of Mathematical Sciences, Ocean University of China, Qingdao, China\\
\email{tlu@stu.ouc.edu.cn, hxiao@ouc.edu.cn, qfang@ouc.edu.cn}}
\date{}
\begin{document}
\maketitle

\begin{abstract}
We consider finite two-person zero-sum mean-payoff games with perfect information, played on a directed graph with Max, Min, and random vertices. We show that, for a fixed number $k$ of random vertices, a deterministic strongly polynomial algorithm computes exact values and uniformly optimal pure stationary strategies. It uses $N^{O(2^k)}$ operations for arena size $N$, independently of reward magnitudes and transition probabilities. The data may be arbitrary real scalars. For rational input and fixed $k$, all intermediate numbers have polynomial bit length. We first solve deterministic discounted games by an affine policy-improvement algorithm, then eliminate random vertices one at a time, and select a discount preserving mean-payoff optimality. One numerical decomposition controls the number of policy updates, sign comparisons, and discount selection.
\end{abstract}

\section{Introduction}
\subsection{Basic concepts}

A BWR-game is played by two players, Max and Min, on a finite directed multigraph $G=(V,E)$. Each vertex has at least one outgoing edge. Write $n=|V|$, $m=|E|$, and $N=2+n+m$. The vertex set is partitioned into $V_{\max}$, $V_{\min}$, and $V_R$. An edge $e$ has tail $i(e)$, head $j(e)$, and reward $r_e\in\R$. At a vertex in $V_{\max}$ or $V_{\min}$, the owner chooses an outgoing edge, whose head is the next vertex. At a random vertex $i$, an outgoing edge $e$ is selected with probability $p_e\geq0$, where $\sum_{i(e)=i}p_e=1$. Set $k=|V_R|$. Zero-probability edges may be retained.

A pure stationary strategy selects one outgoing edge at each vertex of its owner. A pair $\pi=(\sigma,\tau)$ of such strategies is called a policy pair. General strategies may depend on the full observed history and may randomize. For an infinite edge sequence $e_0,e_1,\ldots$, define
$$
\operatorname{MP}=\liminf_{T\to\infty}\frac1T\sum_{h=0}^{T-1}r_{e_h}.
$$
The players optimize $\E[\operatorname{MP}]$. All rewards are bounded because the arena is finite. A pair is uniformly optimal if it is optimal from every initial vertex against all opposing strategies.

When $V_R=\varnothing$, we obtain a deterministic mean-payoff game, also called a BW-game or cyclic game~\cite{GKK,ZP}. If one player has no vertices, we obtain a Markov decision process, and if neither player has vertices, a weighted Markov chain. Simple stochastic reachability games~\cite{Condon} are another special case. Give an absorbing target reward one on its self-loop and assign reward zero elsewhere. The pathwise mean payoff is then the indicator of reaching the target. Pure stationary optimality for perfect-information time-average games is classical~\cite{LL}. Canonical forms for BWR-games were studied in~\cite{BEGMcanonical}, and Blackwell optimality relates the discounted and mean-payoff criteria~\cite{GZ}.

For deterministic games, Gurvich, Karzanov, and Khachiyan~\cite{GKK} gave a potential-reduction algorithm, and Zwick and Paterson~\cite{ZP} obtained a pseudo-polynomial bound by value iteration. Boros, Elbassioni, Gurvich, and Makino~\cite{BEGM} extended pseudo-polynomial solvability to BWR-games with a fixed number of random vertices. For $k\ge1$, integer rewards bounded in absolute value by $R\ge1$, and rational probabilities with common denominator $D$, their bound is $(nDk)^{O(k)}R\operatorname{polylog}(2+R)$. The algorithm uses potential transformations and reduces general games to games whose value is independent of the initial vertex. A convex-programming algorithm gives another pseudo-polynomial bound for fixed $k$~\cite{BEGMconvex}. Allamigeon, Gaubert, Katz, and Skomra~\cite{AGKS} combined value iteration with estimates for bias vectors and value separation. They improved the bounds for solving games with a common value and for finding the vertices of maximal value. Combined with the reduction of~\cite{BEGM}, these estimates improve the fixed-$k$ pseudo-polynomial bound. For fixed $k$ and positive transition probabilities bounded below by an inverse polynomial in the arena size, approximation schemes were obtained in~\cite{BEFGMM}.

The reachability case admits stronger parameter bounds. Gimbert and Horn~\cite{GH} gave algorithms based on permutations of the random vertices. For games with $k\ge1$ fair coin-toss vertices, Ibsen-Jensen and Miltersen~\cite{IJM} obtained the bound $O(k2^k(k\log k+n))$. Auger, Coucheney, and Strozecki~\cite{ACS} gave a randomized algorithm with expected running time $2^{O(k)}$ times a polynomial in the input size, allowing arbitrary outdegrees and rational probability distributions. Andersson and Miltersen~\cite{AM} proved polynomial-time Turing equivalences between solving simple stochastic games and solving stochastic mean-payoff and discounted games with binary numerical data.

For the payoff defined above, a recent preprint of OpenAI~\cite{OpenAIgames} gives a deterministic algorithm for finding the vertices of nonnegative value in a general turn-based stochastic mean-payoff game. Rewards are binary integers and probabilities are rational. The algorithm uses $2^{O((\log(L+2))^2)}$ bit operations, where $L$ is the complete input length.

\subsection{Main result}

Scalar inputs are accessed by exact addition, subtraction, multiplication, division by a nonzero scalar, and comparison. Scalar operations and control operations on explicitly stored bit strings are counted. An algorithm is strongly polynomial if its operation count is bounded by a polynomial in the arena size, and if, on rational input, every intermediate number has bit length polynomial in the complete binary input length.

\begin{theorem}\label{thm:main}
A BWR mean-payoff game with $k$ random vertices can be solved in $N^{O(2^k)}$ operations. The output consists of the exact value at every vertex and a uniformly optimal pure stationary strategy for each player. Rewards and transition probabilities may be arbitrary real scalars. For each fixed $k$, all intermediate numbers on rational input have bit length polynomial in the complete input length.
\end{theorem}

For each fixed $k$, the operation count in Theorem~\ref{thm:main} is independent of the parameters $R$ and $D$ in the pseudo-polynomial bounds of~\cite{BEGM,BEGMconvex,AGKS}.

\subsection{Main ideas of the proof}

Our algorithm first solves deterministic discounted games by policy improvement. Fixing one player's policy gives a one-player game and a guarantee for that player. We improve both policies by one clipped fixed-point computation, reducing the distance between their guarantees by a factor of two. All policy values have a common representation by short integer combinations. This places the possible positive distances in polynomially many numerical scales and bounds the number of rounds.

To eliminate a random vertex, replace it by a terminal of unknown value $x$. Its original value is the unique zero $x^*$ of a decreasing residual. Since the lower-level solver computes affine forms in the rewards along each branch, we can run it symbolically at $x^*$. Each comparison is decided by testing the zero of an affine form and using the monotonicity of the residual. A test requires at most one ordinary lower-level call and one policy evaluation. Repeating this reduction gives the bound $N^{O(2^k)}$.

We then select a discount preserving mean-payoff optimality. For a fixed policy pair, the Bellman residuals are rational functions of the discount. Their numerator coefficients are bounded integer combinations of reward-weighted probability monomials, of which there are polynomially many for fixed $k$. The same numerical decomposition gives an interval on which all residual signs are stable. Solving at one discount in this interval gives a mean-payoff optimal pair, whose values are computed by a linear system.

The use of discounting requires a bound that is uniform as $\beta$ tends to one. Boros et al.~\cite{BEGMdiscount} showed that a sufficient discount may be exponentially close to one even with a single random vertex and one player. Hansen, Miltersen, and Zwick~\cite{HMZ} proved strong polynomiality of strategy iteration for a fixed discount factor, with a bound depending on $1/(1-\beta)$. We therefore require a solver whose operation count is independent of this quantity.

The fixed-point computation uses the extremal-price construction of OpenAI~\cite{OpenAI}. The numerical decomposition follows the simultaneous-approximation principle of Frank and Tardos~\cite{FT}, using rational lattice reduction~\cite{LLL}. Rational linear programming~\cite{GLS} and the Blackwell transfer~\cite{GZ} provide the remaining standard tools.

After the preliminaries in Section~\ref{sec:prelim}, we give the deterministic algorithm and analyze it in Section~\ref{sec:deterministic}. Section~\ref{sec:general} gives the random-vertex reduction, discount selection, and the main algorithm, and concludes with the proof of Theorem~\ref{thm:main}. Throughout, $\log$ has base two and $\ln$ is the natural logarithm.

\section{Preliminaries}\label{sec:prelim}
\subsection{Discounted games and affine computations}\label{sec:affine}

For a common discount factor $0<\beta<1$ and an edge reward vector $d\in\R^m$, the discounted objective is
$$
\operatorname{Disc}_{\beta,d}=\sum_{h\geq0}\beta^h d_{e_h}.
$$
Its value vector $v$ is the unique fixed point of the contraction
$$
(T_{\beta,d}v)_i=
\begin{cases}
\max_{i(e)=i}(d_e+\beta v_{j(e)}),&i\in V_{\max},\\
\min_{i(e)=i}(d_e+\beta v_{j(e)}),&i\in V_{\min},\\
\sum_{i(e)=i}p_e(d_e+\beta v_{j(e)}),&i\in V_R.
\end{cases}
$$
Choosing a tight edge at every controlled vertex gives a uniformly optimal pure stationary pair for the discounted game.

The discounted solver will satisfy an additional property. For fixed transition probabilities and fixed $\beta$, its scalar computation has two kinds of registers. Coefficient registers are independent of the reward vector along a fixed execution branch. Reward registers contain affine forms in the edge rewards. Reward registers may be added or subtracted, multiplied or divided by coefficient registers, and compared. A branch can depend on any such comparison. Coefficient calculations may depend on the discrete branch selected so far. This structure is an implementation property, including every intermediate calculation and comparison.

The coefficients of a reward register can be propagated explicitly by carrying its vector of $m+1$ affine coefficients. The construction below provides a uniform bit bound on these coefficients, independently of the reward input.

The policy solvers return $(\sigma,\tau)$. The routine \textsc{EvaluatePair} returns mean-payoff values when no discount is supplied, and unnormalized discounted values when a final argument $\beta\in(0,1)$ is supplied. In each call, $G$ contains the arena, ownership, and probabilities; rewards are a separate argument. Fix deterministic tie-breaking and fixed vertex and edge orders.

\subsection{Markov chains and policy evaluation}

Once both policies are fixed, the game becomes a finite Markov chain. Its discounted and mean-payoff values can be computed by Gaussian elimination. For a pair with transition matrix $P$ and expected reward vector $\bar r$, write $A=I-P$. Discounted values solve $(I-\beta P)w=\bar r$. Mean-payoff values are the $g$ component of
$$
Ag=0,\qquad g+Ah=\bar r.
$$
The auxiliary vector $h$ need not be unique.

\begin{lemma}\label{lem:evaluation}
For a fixed policy pair, \textup{\textsc{EvaluatePair}} computes its mean-payoff vector, or its discounted value vector when $0<\beta<1$ is supplied, by a linear system of order at most $2n$. Its operation count is polynomial in $N$. For fixed probabilities and any supplied discount, the computation has the affine register structure, with coefficient lengths polynomial in $N$ and the rational encoding length of these data, independently of rewards. All rational intermediates have polynomial bit length in the complete input.
\end{lemma}
\begin{proof}
For discounted values, $I-\beta P$ is nonsingular because $P$ is stochastic and $\beta<1$.

The mean-payoff system is consistent and has a unique $g$. Indeed, if $x=Ay$ and $Ax=0$, then $P^t y=y-tx$ for every integer $t\ge0$. The bound $\norm{P^t y}_\infty\le\norm{y}_\infty$ implies $x=0$. Rank-nullity therefore gives $\R^n=\ker A\oplus\operatorname{im}A$, which uniquely decomposes $\bar r$ as $g+Ah$ with $Ag=0$.

This $g$ is the mean-payoff vector. Since $Pg=g$ and $\bar r=g+(I-P)h$,
$$
\frac1T\sum_{t=0}^{T-1}P^t\bar r
=g+\frac{h-P^T h}{T}\longrightarrow g.
$$
The reward average of a finite Markov chain converges almost surely. Bounded convergence identifies its expected limit, and hence its expected liminf, with the displayed limit.

Gaussian elimination handles both systems, setting free variables to zero in the second. Pivots and matrix calculations depend only on $P,\beta$; right-hand-side calculations are linear in rewards. Standard minor bounds give polynomial coefficient lengths and rational intermediate lengths.
\end{proof}

\subsection{Deterministic one-player games}

Post and Ye~\cite{PY} gave strongly polynomial algorithms for discounted deterministic Markov decision processes with arbitrary discounts. For a common discount, we use a direct computation by paths and closed walks to obtain the required affine implementation.

\begin{proposition}\label{prop:mdp}
On a deterministic arena with choices for only one player, exact discounted values and a uniformly optimal stationary policy can be computed in polynomially many operations uniformly over $0<\beta<1$. The computation has the affine register structure. Its coefficient lengths are polynomial in $N$ and the bit length of rational $\beta$, independently of the rewards.
\end{proposition}
\begin{proof}
Consider maximization. Compute by finite-length dynamic programming the largest discounted reward $D_a(s,v)$ of a walk of exactly $a$ steps from $s$ to $v$, for all endpoints and $0\le a\le n$. Unattainable endpoint pairs are stored as discrete absences. The recurrence is
$$
D_{a+1}(s,v)=\max_{e:u\to v}\bigl(D_a(s,u)+\beta^a d_e\bigr).
$$
For each starting vertex $s$, maximize
$$
D_a(s,v)+\frac{\beta^aD_b(v,v)}{1-\beta^b}
$$
over attainable choices $0\le a<n$, $1\le b\le n$, and $v\in V$. Each candidate is the payoff of a feasible prefix followed by repetitions of a closed walk. A stationary discounted-optimal policy exists by the Bellman contraction, and its prefix and cycle occur among these candidates. Thus their maximum equals the value. A Bellman-tight action at every vertex gives a uniformly optimal policy.

There are polynomially many states, lengths, and candidates. Each chosen walk has at most $n$ edges. Along a branch all computed quantities are linear in rewards, with coefficients formed from $\beta^a$ and $1/(1-\beta^b)$ for $a,b\le n$. Their rational lengths are polynomial in $N$ and the encoding length of $\beta$. Minimization is obtained by negating rewards.
\end{proof}

\subsection{A shared separation and compression routine}\label{sec:separation}

The routines \textsc{Separate} and \textsc{CompressSigns} are two outputs of one decomposition. We give its linear-access form using rational LLL reduction~\cite{LLL}, following the simultaneous-approximation principle behind Frank--Tardos compression~\cite{FT}.

\begin{lemma}\label{lem:separation}\label{cor:compression}
For $w\in\R^s$, $s\ge1$, and an integer $B\ge1$, one can compute $\delta>0$ and a short integer vector $\widehat w$ such that every $a\in\Z^s$ with $\norm{a}_1\le B$ satisfies
$$
\sgn(a^\top\widehat w)=\sgn(a^\top w),\qquad
a^\top w\ne0\ \Longrightarrow\ |a^\top w|\ge\delta.
$$
Moreover, all nonzero magnitudes $|a^\top w|$ with $\norm{a}_1\le B$ lie in at most $s$ intervals $[S,AS]$, where $\log A=\poly(s,\log(2B))$ is independent of $w$. The operation count and output length of $\widehat w$ are polynomial in $s,\log(2B)$, independently of scalar magnitudes. Accesses to $w$ are signs of rational linear forms with polynomial coefficient length, and arithmetic involving $w$ uses such forms. On rational input all intermediate numbers have polynomial length in the complete input and $\log(2B)$.
\end{lemma}
\begin{proof}
For $w=0$, return $(1,0)$. Otherwise set $\varepsilon=1/(2B)$ and $Q=2^{(s+1)^2}(8/\varepsilon)^{s+1}$. Starting with $v=w$, choose $X=\norm{v}_\infty>0$ as a signed coordinate and find a short rational vector $u$ with $\norm{u-v/X}_\infty\le\varepsilon/(4Q)$ by binary search. Every comparison is $v_i-aX\gtreqless0$ with short rational $a$, so no division by $X$ is performed on the real input.

The lattice $\{(qu-z,q/Q):z\in\Z^s,q\in\Z\}$ has determinant $1/Q$. LLL supplies a nonzero vector of norm at most $2^{s/2}Q^{-1/(s+1)}<\varepsilon/8$. Its $q$ is nonzero; choose its sign so $1\le q\le Q$. Then $\norm{qv/X-z}_\infty<\varepsilon$. Put $\lambda=X/q$ and replace $v$ by $v-\lambda z$. A previously zero coordinate stays zero, while a coordinate attaining $\pm X$ vanishes because its integer digit is $\pm q$. Thus at most $p\le s$ steps give
$$
w=\sum_{j=1}^p\lambda_jz_j,\qquad
\left\|w-\sum_{h=1}^j\lambda_hz_h\right\|_\infty\le\varepsilon\lambda_j,
\qquad \norm{z_j}_\infty\le Q+1.
$$
Set $A=2B(Q+1)+1$. For an allowed form, its first nonzero integer digit $a^\top z_j$ has magnitude between one and $B(Q+1)$. The remaining tail has magnitude at most $B\varepsilon\lambda_j=\lambda_j/2$, so the form has that digit's sign and
$$
\lambda_j/2\le |a^\top w|\le A\lambda_j/2.
$$
These $p\le s$ intervals give the magnitude claim, with $\log A=\poly(s,\log(2B))$. If all digits vanish, the form is zero. Return $\delta=\tfrac12\min_j\lambda_j$ and $\widehat w=\sum_{j=1}^p A^{p-j}z_j$.
The same first digit determines the sign on $\widehat w$, since its later integer-place tail has magnitude less than half the leading place.

All lattices have short rational bases and polynomial dimension. Every scale and remainder is a rational linear form in $w$; in one step coefficient denominators grow by at most $Q$ and integer digits by at most $Q+1$. There are at most $s$ steps, so coefficient lengths remain polynomial. LLL and digit assembly use only short bit strings.
\end{proof}

\section{Deterministic discounted games}\label{sec:deterministic}

We first solve discounted games with $V_R=\varnothing$. The random-vertex reduction will use both the uniform operation bound and the affine implementation given below.

\begin{theorem}\label{thm:base}
There is a uniform deterministic algorithm $A_0$ returning a uniformly optimal pure stationary pair for deterministic two-player discounted games in $\poly(N)$ operations, uniformly over $0<\beta<1$ and real rewards. It has the affine register structure. On rational $\beta$, all coefficient registers and all affine coefficients have length polynomial in $N$ and the length of $\beta$, uniformly over rewards and execution branches.
\end{theorem}

\subsection{Description of the policy-improvement algorithm}

We maintain a policy for each player and recompute their guarantees after every update. Algorithm~\ref{alg:deterministic} gives the $k=0$ case of the recursive solver. Put $c(y)_e=d_e+\beta y_{j(e)}-y_{i(e)}$. One clipped fixed point determines the new actions. An action is changed only where the corresponding bound improves the old guarantee.

\begin{algorithm}[H]
\caption{Deterministic discounted policy improvement}\label{alg:deterministic}
\small
\begin{algorithmic}[1]
\Require A deterministic arena $G$, common discount $0<\beta<1$, and rewards $d$
\Ensure Uniformly optimal discounted policies $\sigma,\tau$
\State Choose any policies $\sigma,\tau$
\Loop
  \State $(L,U)\gets\Call{PolicyGuarantees}{G,\beta,d,\sigma,\tau}$ and $W\gets\norm{U-L}_\infty$
  \If{$W=0$}
    \State \Return $(\sigma,\tau)$
  \EndIf
  \State $y\gets(L+U)/2$ and $H\gets W/4$
  \State $(u,\xi)\gets\Call{ClippedSolve}{G,c(y),H}$
  \State $z^-\gets y+u-H\one$ and $z^+\gets y+u+H\one$
  \State Set $\sigma(i)\gets\xi(i)$ at Max vertices with $z_i^->L_i$; retain the other actions
  \State Set $\tau(i)\gets\xi(i)$ at Min vertices with $z_i^+<U_i$; retain the other actions
\EndLoop
\end{algorithmic}
\end{algorithm}

\textsc{PolicyGuarantees} returns the exact Min MDP value $L$ with $\sigma$ fixed and the exact Max MDP value $U$ with $\tau$ fixed, using Proposition~\ref{prop:mdp}.

The call $\textsc{ClippedSolve}(G,c,H)$, for $H>0$, returns a fixed point $u$ of
\begin{equation}\label{eq:clipped}
u_i=\clip_{[-H,H]}\left(\operatorname{ext}_{e:i\to j}(c_e+u_j)\right),
\end{equation}
and an attaining edge $\xi(i)$ at every controlled vertex for the extremum inside the clipping operation. Lemma~\ref{lem:clipped-affine} and Algorithm~\ref{alg:clipped} supply this routine and its affine implementation.

\subsection{Correctness of the policy update}

Let $v$ be the discounted value vector. The guarantees of the current policies satisfy $L\le v\le U$. We show that the update preserves these inequalities and reduces the width by a factor of two.

\begin{lemma}\label{lem:clipped-update}
Let $L,U$ be the guarantees of the current Max and Min policies, with $W=\norm{U-L}_\infty>0$. Set $y=(L+U)/2$ and $H=W/4$, and obtain $(u,\xi)$ from \textup{\textsc{ClippedSolve}} with offsets $c(y)$. Define $z^\pm=y+u\pm H\one$ and update the two policies as in Algorithm~\ref{alg:deterministic}. Their exact new guarantees $L',U'$ satisfy, componentwise,
$$
L'\ge\max\{L,z^-\},\qquad U'\le\min\{U,z^+\}.
$$
In particular, $L\le L'\le v\le U'\le U$ and $\norm{U'-L'}_\infty\le W/2$.
\end{lemma}
\begin{proof}
The vector $q=\max\{L,z^-\}$ is a subsolution with the updated Max policy fixed. Where $q_i=L_i$, the old Max action, or every Min action, satisfies $d_e+\beta q_j\ge L_i$ because $q\ge L$. Where $q_i=z_i^->L_i$, one has $u_i>-H$, since $u_i=-H$ would imply $z_i^-=y_i-2H\le L_i$. The clipped equation then gives $c(y)_e+u_j\ge u_i$ for the selected Max action and every Min action. Hence
$$
d_e+\beta q_j-q_i
\ge d_e+\beta z_j^--z_i^-
=c(y)_e+u_j-u_i+(1-\beta)(H-u_j)\ge0.
$$
Iteration of this monotone contraction gives $L'\ge q$. Negating rewards and exchanging the players gives $U'\le\min\{U,z^+\}$. Policy guarantees bracket $v$, and $z_i^+-z_i^-=2H=W/2$ bounds the new width.
\end{proof}

\subsection{Auxiliary fixed-point computations}\label{sec:shortdata}

The clipped fixed-point routine first compresses the offsets by Lemma~\ref{lem:separation}. It then approximates a fixed point and recovers an exact solution on one affine branch. We find a basis by rational linear programming~\cite{GLS} on the compressed data and transfer it to the original offsets.

\subsubsection{Extremal-price search}\label{app:prices}

The following construction is adapted from OpenAI~\cite[Theorem~2 and Section~6]{OpenAI}. Its dimension is $d$; the arena-size parameter remains $N$. For a signed coordinate $s=(i,h)$ write $y_s=hy_i$.

\begin{theorem}[Extremal prices]\label{oa:thm:prices}
Let $d,R\ge1$ be integers and let $\Slots=[d]\times\{-1,+1\}$.
There is a deterministic algorithm that queries price vectors
$p\in\Q^d$, receives a label vector $z\in\{-1,+1\}^d$ for each
query, and returns $x\in\Q^d$ with the following guarantee.

A hypothesis consists of a nonempty set $S\subseteq\Slots$ and
integer demands $a_s\in[-R,R]$ for $s\in S$. Its surplus at
$s=(i,h)$ is $d_s(p)=a_s-hp_i$. Call the hypothesis
\emph{consistent} if at every query and answer $(p,z)$,
$$
   s=(i,h)\in S,\qquad
   d_s(p)=\max_{u\in S}d_u(p)\ge0
   \quad\Longrightarrow\quad z_i=h.
$$
For each consistent hypothesis the output meets all its demands,
that is, $h x_i\ge a_s$ for every $s=(i,h)\in S$.
The bit time, including all query and answer bits, is polynomial in
$d$ and $\log(R+1)$.
\end{theorem}
\par\ifdim\dimexpr\pagegoal-\pagetotal\relax<6\baselineskip\newpage\fi
\begin{proof}\leavevmode\par
\paragraph{Market recursion.}

Choose $\delta=1/5$, $\gamma=3/16$, $\eta=1/16$, and let $D$ be the least positive odd integer with $(5/4)^D>2d$. The expert pool $E$ consists of all ordered tuples of four signed coordinates with integer demand guesses in $[-R,R]$, together with a threshold description. The latter is either zero or a level $j\in\{0,\ldots,D-1\}$ and a signed pivot coordinate with its own demand guess. Repetitions, including different guesses on the same coordinate, are allowed. Its size $|E|=[2d(2R+1)]^4[1+2dD(2R+1)]$ is polynomial in $d,R$ and can be computed without enumeration.

At level $\ell$, an expert with zero threshold uses $\tau=0$. An expert specifying $j<\ell$ uses $\tau=b-y_s^j$, where $b,s$ are its pivot guess and coordinate and $y^j$ is the history proposal at level $j$. Experts with $j\ge\ell$ are inactive. An active expert has four options of utility $b-\tau-y_s$, one per guessed coordinate, and all experts have an outside option of utility zero. The estimate $\tau$ is not clipped.

For positive wealths $w_e$, let $U_e(y)$ be the maximum option utility and minimize $\Phi(y)=\sum_e w_eU_e(y)$. Given feedback $v\in\{\pm1\}^d$, let $g_e$ be the least payoff among maximizing options, paying $v_s$ to a coordinate option and zero outside. At a minimizer,
\begin{equation}\label{oa:eq:market}
\sum_e w_eg_e=-\Phi'(y;v)\le0,
\end{equation}
because the directional derivative of a maximum is its largest active slope. Hence a market update by $1+\eta g_e$ preserves positive wealth and cannot increase its total.

Set
\begin{equation}\label{oa:eq:probepars}
c=2^{-14},\qquad r=2^{19}+1,
\end{equation}
and $r_0=r+3072\lceil\log(|E|+1)\rceil$. The root starts with wealth one per expert. A level-$D$ leaf returns the coordinatewise median of its $D$ history proposals and its input wealth. Every internal node makes $r$ probes, or $r_0$ at the root. Each probe consists of three operations.
\begin{enumerate}
\item Choose a market minimizer $y$ using wealth $(1-c)w$.
\item Call a fresh child with wealth $cw$ and $y$ appended to the history, obtaining prices $x$ and wealth $w'$.
\item Query labels $z$ at $x$, set
\begin{equation}\label{oa:eq:feedback}
v_i=\begin{cases}\sgn(x_i-y_i),&x_i\ne y_i,\\z_i,&x_i=y_i,\end{cases}
\end{equation}
and replace each wealth by
\begin{equation}\label{oa:eq:invest}
w_e\gets(1-c)w_e(1+\eta g_e)+w'_e.
\end{equation}
\end{enumerate}
The node retains these changes between probes and returns the median child prices and final wealth. Output the root prices. Equation~\eqref{oa:eq:market} and induction on the tree show that every node returns total wealth at most its input total.

\paragraph{Thresholds and feedback.}\label{oa:sec:amplify}

Fix the complete run and a consistent hypothesis $(S,a)$. For the analysis, assign thresholds and distributions to the nodes. A node with threshold $t\ge0$ succeeds if all its returned surpluses are at most $t$. Start at the root with $t=0$ and the uniform distribution $m$ on $S$.

At a proposal $y$, sort the mass of $m$ by nonincreasing surplus. A probe is light if $m\{d_s(y)>t\}\le\delta$; assign its child $t'=t$. Otherwise it is heavy. Select a pivot meeting the mass band $[\gamma,\delta)$ in positive length and set $t'$ to its surplus. Then $t'>t$ and
\begin{equation}\label{oa:eq:band}
m\{d_s(y)\ge t'\}\ge\gamma,\qquad m\{d_s(y)>t'\}\le\delta.
\end{equation}
For either type, form $m'$ by multiplying the first $\delta$ of sorted mass by four, partially boosting a coordinate crossing the cutoff, and dividing by $1+3\delta$. Every coordinate with $d_s(y)>t'$ is fully boosted. The distribution $m'$ is independent of the allowed pivot choice. Thresholds are recorded as zero or as a pivot and its proposal level.

A successful child gives
\begin{equation}\label{oa:eq:correct}
d_s(y)\ge t'\quad\Longrightarrow\quad v_s=1.
\end{equation}
Indeed, $d_s(x)\le t'\le d_s(y)$ implies $x_s\ge y_s$. Strict inequality gives the sign directly. Equality makes $s$ a maximum-surplus coordinate at $x$ with surplus $t'\ge0$, so consistency supplies the same sign.

Every leaf succeeds under any such threshold assignment.

For a contradiction, suppose some coordinate exceeds the final threshold at a majority of the $D$ history proposals. Thresholds never decrease, so that coordinate was fully boosted at each of those proposals. Its final mass is at least $(2d)^{-1}4^{D/2}/(1+3\delta)^D=(2d)^{-1}(5/4)^D>1$, a contradiction.

At a node let $\nu$ sample four coordinates independently from $m$, use their true demands, and match the current threshold description. These experts have $\tau=t$.
If a probe's child succeeds, $\sum_e\nu_e\ln(1+\eta g_e)\ge0$. In a heavy probe this average is at least $\epsilon=1/256$.

In a light probe, every maximizing option receives nonnegative feedback by~\eqref{oa:eq:correct}. In a heavy probe, a tuple hits $\{d_s(y)\ge t'\}$ with probability at least $1-(1-\gamma)^4$. On a hit all maximizing options receive payoff one, since $t'>t$ excludes the outside option. Since other gains are at least $-1$, the mean gain exceeds $1/8$. The inequality $\ln(1+\eta g)\ge\eta g-\eta^2$ gives the bound $\eta/8-\eta^2=\epsilon$.

\paragraph{Wealth bound.}

At level $\ell$, augment $\nu$ by reserving mass $2^{\ell-j}$ for each future pivot level $j\in\{\ell,\ldots,D-1\}$. For this reserve, sample the four coordinates and the pivot independently from $m$, with true demands. Denote the resulting expert weights by $\pi$. Their total is at most three, and the reserve experts have zero market gain at this node because they are inactive.

Write $\mu_e$ for an expert's returned wealth divided by its input wealth, and put $L_e=\ln(1+(c/2)\mu_e)\ge0$. These quantities depend on the fixed run, not on proof thresholds. A probe has multiplier $(1-c)(1+\eta g_e)+c\mu'_e$, which is at least $(1-c)(1+\eta g_e)(1+(c/2)\mu'_e)$ because $(1-c)(1+\eta g_e)<2$. Consequently
\begin{equation}\label{oa:eq:bonus}
L_e\ge-\ln(2/c)+\sum_{\text{probes}}\bigl(\ln(1-c)+\ln(1+\eta g_e)+L'_e\bigr).
\end{equation}

For each run and consistent hypothesis, allowed pivots can be chosen so that every parent--child probe satisfies $\sum_e\pi'_eL'_e\le C\sum_e\pi_eL'_e$, where $C=80(5/2)^4=3125$.

Choose pivots down the tree. In a heavy probe, a uniform point of $[\gamma,\delta)$ selects a pivot with probabilities at most $80m_s$. Since $m'\le(5/2)m$, the child's current-threshold distribution is dominated in expectation by $C$ times the parent's reserve for $j=\ell$. In a light probe it is dominated by $(5/2)^4\nu$. Each future reserve costs at most $2(5/2)^5<C$ relative to its parent reserve, accounting for its doubled mass and five sampled coordinates. These are distinct expert classes, so $\EE\pi'\le C\pi$ componentwise. The bonuses $L'_e$ are fixed and nonnegative; averaging therefore gives an allowed pivot satisfying the asserted inequality. Later choices do not change those bonuses.

With these pivots, a failed internal node with $T$ probes satisfies
\begin{equation}\label{oa:eq:charge}
\sum_e\pi_eL_e\ge\sigma T,\qquad \sigma=\epsilon/4=1/1024.
\end{equation}

We induct upwards. A successful child contributes at least zero to $\sum_e\pi_e(\ln(1+\eta g_e)+L'_e)$, and at least $\epsilon$ if heavy, by the successful-child bound above. A failed child is internal because every leaf succeeds. Induction and the chosen pivots then give a contribution at least $-2\eta+\sigma r/C\ge\epsilon$; only the unit-mass current-threshold experts can have negative market gains. If the parent fails, a majority of its probes are heavy or have failed children, by the median rule. Using~\eqref{oa:eq:bonus}, $\sum_e\pi_e\le3$, and $\ln(1-c)\ge-2c$, we obtain
$$
\sum_e\pi_eL_e\ge T\epsilon/2-6cT-3\ln(2/c)\ge\sigma T.
$$
The last inequality follows from $T\ge r$, $6c<\epsilon/8$, and $3\ln(2/c)<r\epsilon/8$.

At the root, total wealth is at most $|E|$ and input wealth is one per expert, so $L_e\le\log(|E|+1)$. But $\sigma r_0>3\log(|E|+1)$, ruling out root failure by~\eqref{oa:eq:charge}. This holds separately for every consistent hypothesis, proving the demand guarantee.

\paragraph{Bit implementation.}\label{oa:sec:implement-pr}

There are $O(r_0r^{D-1})=O(d^{60}\log(dR+1))$ probes, since $\log_{5/4}r<60$. All wealths have a common dyadic denominator gaining only constantly many places per probe, and total wealth is bounded by $|E|$. Thus their bit lengths are polynomial.

Market prices also stay short. Choose a vertex of the optimal epigraph face. Such a bounded face exists because zero-threshold experts repeating a single signed coordinate with guess zero make $\Phi$ coercive. The prices at a vertex are determined by active option ties; otherwise a nonzero direction preserving all ties gives an optimal line segment. Slot--slot ties have the form $y_i=\pm y_j+B$, with integer $|B|\le2R$. Along a spanning tree in each connected component, all prices equal a signed reference price plus an integer of magnitude $O(dR)$. Each component is pinned either by a signed cycle fixing twice its reference to an integer, or by an outside-option tie fixing a coordinate to an integer or a signed history price plus an integer of magnitude at most $2R$. Induction on the history depth gives half-integral prices of magnitude $O(DdR)$. Medians preserve both bounds.

The exponentially large expert pool is stored implicitly. Group by the four signed coordinates, the threshold level or zero description, and the pivot coordinate, excluding all demand guesses. This leaves at most five integer variables per group. Every option comparison bounds a single guess or a difference of two guesses. Maintain partitions of their integer ranges at all comparison thresholds encountered at actual proposals, separating ties. A cell selects one interval per range. Each range gains only constantly many cuts per probe and there are constantly many ranges per group, so the number of cells is polynomial. Experts in a cell have common maximizing option positions, gains, and per-expert wealth. Refinement copies that wealth, including the records of active ancestor probes.

Cell counts and first moments are computable without enumeration. Eliminate guess variables successively. Each variable ranges from a maximum of constant or shifted-variable lower bounds to a minimum of such upper bounds. Subdivide by which bounds attain these extrema and require the lower bound not to exceed the upper. These conditions remain interval bounds on guesses and their differences. Summing a polynomial over the resulting integer interval gives a polynomial by the power-sum formulas, also for negative endpoints. There are at most five eliminations, of constant branching and degree, so counts and coordinate sums have polynomial bit cost.

Finally, construct a separator for the epigraph of $\Phi$. At rational test prices, temporarily refine cells by the option comparisons, choose a maximizing option in each cell, and use its count, first moments, and wealth to form an affine function attaining $\Phi$ there and lying below it everywhere. This separates a point whose proposed height is too small. All epigraph inequalities have polynomial bit length: they are weighted sums of option utilities with short history prices and a common short wealth denominator, and $\log|E|$ is polynomial. The separator has polynomial bit cost, including the test-price bits. Rational separation--optimization and successive coordinate minimization produce the lexicographically first optimal vertex in polynomial time~\cite[Theorem~(6.4.9) and Lemma~(6.5.1)]{GLS}. The bounded optimal face and the vertex price bound justify these successive calls. Updating the persistent cells at that proposal implements its market gains. Labels, medians, and all remaining operations have the same bit bounds.
\end{proof}

\subsubsection{Approximation and affine recovery}

\begin{lemma}\label{lem:approximate}
Let $f:[-1,1]^d\to[-1,1]^d$ be nonexpansive in the maximum norm. Given a dyadic evaluation oracle for $f$, a dyadic point $x$ with $\norm{f(x)-x}_\infty\le 2^{-q}$ can be found in bit time polynomial in $d,q$, including the lengths of all queries and answers. The oracle is asked for only polynomially many bits of accuracy.
\end{lemma}
\begin{proof}
We first treat a contraction with Lipschitz constant at most $1-\gamma$, where $\gamma=2^{-b}$, and approximate its fixed point $v$ to distance $\epsilon=2^{-q}$. Set $\delta=\gamma\epsilon/16$ and $R=2/\delta$, an integer. Extend $f$ to all of $\R^d$ by clipping its argument to the cube. Apply the extremal-price algorithm of Theorem~\ref{oa:thm:prices} with dimension $d$ and demand bound $R$. At query prices $p$, put $x=\delta p$. Round its clipped argument to the nearest $\delta$-grid point and request $b+q+4$ bits from the oracle. Subtract the original $x$ when forming residual signs, resolving zeros deterministically. Relative to the extended map $f(x)-x$, the evaluation error is at most $2\delta$.

For the analysis, assign the hidden demands
$$
a_{i,h}=\left\lfloor\frac{hv_i-\epsilon/2}{\delta}\right\rfloor
\qquad (i\in[d],\ h\in\{-1,1\}).
$$
They belong to $[-R,R]$. If $(i,h)$ maximizes their surplus at $p$ and this surplus is nonnegative, write $u=h(v_i-x_i)$. The floor inequalities give $u\ge\epsilon/2$ and $\norm{v-x}_\infty\le u+\delta$. Consequently
$$
h(f_i(x)-x_i)\ge u-(1-\gamma)(u+\delta)
\ge\gamma\epsilon/2-\delta=7\delta.
$$
The returned label is therefore $h$. These demands form a consistent hypothesis for Theorem~\ref{oa:thm:prices}. Its output meets all demands, giving $\norm{x-v}_\infty\le\epsilon/2+\delta$. Clipping and rounding once more gives a dyadic point within $\epsilon$ of $v$. The bit bound follows from $\log R=b+q+5$ and the bound on all prices in that theorem.

For nonexpansive $f$, apply this contraction construction to $g=(1-\epsilon/2)f$ with target distance $\epsilon/4$. At its fixed point $v_g$, $\norm{v_g-f(v_g)}_\infty\le\epsilon/2$. Changing the argument by $\epsilon/4$ changes this residual by at most $\epsilon/2$. Thus the returned point has residual at most $\epsilon$. An $a$-bit evaluation of $g$ is obtained by requesting $a+1$ bits from $f$, multiplying by $1-\epsilon/2$, and rounding to $a$ bits. The required contraction gap and target accuracy have logarithms $O(q)$.
\end{proof}

Fixing extremizing edges and clipping branches in~\eqref{eq:clipped} gives a bounded polytope $Au\le Bw$, where $w=(c,H)$ and $A,B$ are short integer matrices. A vertex basis $S$ yields $u=A_S^{-1}B_Sw$. Algorithm~\ref{alg:clipped} selects such a basis on compressed data and transfers it to the original offsets.

\begin{algorithm}[H]
\caption{Affine recovery of a clipped fixed point}\label{alg:clipped}
\small
\begin{algorithmic}[1]
\Require A deterministic arena $G$, offsets $c$, and height $H>0$
\Ensure A clipped fixed point $u$ and attaining edges $\xi$
\State $w\gets(c,H)$ and $(\widehat c,\widehat H)\gets\Call{CompressSigns}{w,2^{16N^3}}$
\State Form $f_i(x)=\clip_{[-1,1]}(\operatorname{ext}_{e:i\to j}(\widehat c_e/\widehat H+x_j))$
\State Let $L\ge2$ be its encoding length and put $p(L)=64(L+1)^4$
\State $\widetilde x\gets\Call{ApproximateFixedPoint}{f,2^{-p(L)-1}}$
\State Choose an extremum-and-clipping branch $\chi$ realizing $f(\widetilde x)$
\State Use rational LP to find a vertex basis $S$ of $\chi$'s exact fixed-point polytope
\State $u\gets A_S^{-1}B_Sw$ and let $\xi$ be the attaining edges specified by $\chi$
\State \Return $(u,\xi)$
\end{algorithmic}
\end{algorithm}

\begin{lemma}\label{lem:clipped-affine}
For a deterministic arena, real offsets $c_e$, and $H>0$, a solution of the clipped equation~\eqref{eq:clipped} and extremizing actions at every vertex can be computed in $\poly(N)$ operations. The algorithm accesses $w=(c,H)$ through rational linear combinations and signs of rational linear forms. On each execution branch its output is a rational linear map of $w$. Every coefficient in these combinations and maps has bit length $\poly(N)$.
\end{lemma}
\begin{proof}
Substitution into the remaining inequalities, followed by multiplication by a constant positive denominator, reduces feasibility to signs of integer linear forms on $w$. The adjugate formula bounds all their coefficient $1$-norms by $M=2^{16N^3}$, uniformly over branches and bases. The same bound covers the test $H>0$.

Lemma~\ref{cor:compression} gives short integer data $\widehat w=(\widehat c,\widehat H)$ preserving these signs, with $\widehat H>0$. The normalized clipped map in Algorithm~\ref{alg:clipped} is nonexpansive and has encoding length $L=\poly(N)$. Exact evaluation at dyadic arguments has polynomial bit cost; rounding supplies the oracle of Lemma~\ref{lem:approximate}. Its affine cells have a uniform residual gap. On any cell, introduce the gate values and minimize $e$ subject to $0\le e\le2$ and $-e\le f_i(x)-x_i\le e$ in the cube. This is a bounded rational polytope. Clearing denominators gives matrices of order $O(L)$ with $O(L^2)$-bit entries. Determinant bounds imply that a positive minimum is at least $2^{-p(L)}$, where $p(L)=64(L+1)^4$ suffices.

Lemma~\ref{lem:approximate} supplies a point with residual at most $2^{-p(L)-1}$. Exact gate evaluation selects a realizing cell, whose minimum residual must therefore be zero. Rational LP finds a vertex of its exact fixed-point polytope in the $\widehat u$ coordinates and extracts a basis $S$. Sign preservation transfers its feasibility to the original data, yielding $u=A_S^{-1}B_Sw$ and the attaining edges.

Compression accesses $w$ only through short rational linear sign queries. Approximation and LP then use short compressed data, and recovery is multiplication by a short rational matrix. All accesses to the original offsets are consequently affine, with polynomial coefficient lengths.
\end{proof}

\subsection{Termination and running time}

\begin{proof}[Proof of Theorem~\ref{thm:base}]
All possible guarantee widths use a common finite-dimensional representation. A fixed pair reaches its first repeated vertex in at most $n$ steps. Summing the periodic tail geometrically expresses its value at any vertex as an integer linear combination of at most $n$ entries of
$$
w=\left((\beta^h d_e)_{e\in E,\,0\le h<n},\left(\frac{\beta^h d_e}{1-\beta^b}\right)_{e\in E,\,0\le h<n,\,1\le b\le n}\right).
$$
This vector has dimension $s=mn(n+1)$. By Proposition~\ref{prop:mdp}, each guarantee is a policy-pair value vector. Every positive width is one coordinate difference of two such vectors, hence $|a^\top w|$ for an integer $a$ with $\norm{a}_1\le2n$. Lemma~\ref{lem:separation} places all these magnitudes in at most $s$ intervals $[S,AS]$, with $\log A=\poly(N)$ independent of rewards and discount. Each interval contains at most $1+\lceil\log A\rceil$ terms of a halving sequence. Lemma~\ref{lem:clipped-update} therefore forces zero width, and thus optimality, within polynomially many rounds. Each round has polynomial cost by Proposition~\ref{prop:mdp} and Lemma~\ref{lem:clipped-affine}.

One round is affine with uniformly short coefficients. Its MDP evaluations use powers $\beta^a$ and reciprocals $1/(1-\beta^b)$ for $a,b\le n$. Widths, centers, offsets, and radii are affine on each branch; the clipped solve uses a short rational matrix and short linear sign queries. Action switches compare affine bounds. Only discrete policies persist between rounds, so recomputing guarantees from the original rewards prevents coefficient accumulation. Explicitly carrying coefficients has polynomial overhead, and evaluating them at rational rewards gives the numerical bit bound.
\end{proof}

\section{General BWR-games}\label{sec:general}

We extend the discounted solver by eliminating one random vertex at a time. After selecting a common discount, we apply this solver and evaluate the resulting policies for mean payoff.

\subsection{Preliminary results and basic lemmas}
\subsubsection{Parametrized discounted games}\label{sec:elimination}

Consider a family of games in which one terminal reward varies and all other data remain fixed.

Fix $\beta\in(0,1)$ and $c\in V_R$. Let $G_c$ replace all outgoing edges of $c$ by one self-loop assigned to either player. Its rewards $d^{(x)}$ retain $d$ elsewhere and give the loop reward $(1-\beta)x$. This removes one random vertex without increasing the arena. Write $G_c(x)=(G_c,\beta,d^{(x)})$ and $U(x)$ for its value vector, with $U_c(x)=x$.

Using the original outgoing data at $c$, define
$$
g(x)=\sum_{i(e)=c}p_e\bigl(d_e+\beta U_{j(e)}(x)\bigr)-x.
$$
Write $x^*$ for the original discounted value at $c$.

\begin{lemma}\label{lem:root}
For every vertex $j$ and every $y>x$,
$$
0\leq U_j(y)-U_j(x)\leq y-x.
$$
The function $g$ defined above is continuous and satisfies
$$
g(y)-g(x)\leq -(1-\beta)(y-x).
$$
It has a unique zero $x^*$, which is the value at $c$ in the original discounted game. The vector $U(x^*)$ is the original value vector.
\end{lemma}

\begin{proof}
Fix any pair of general strategies and let $\tau_c$ be the first hitting time of $c$, with $\beta^{\infty}=0$. Changing the terminal value from $x$ to $y$ changes the expected payoff from $j$ by $\E_j[\beta^{\tau_c}](y-x)$. This number lies in $[0,y-x]$. Taking the players' extrema preserves the two bounds. In particular, each $U_j$ is continuous. Substituting the bounds in the definition of $g$ gives its strict decrease.

Let $v$ be the fixed point of the original discounted Bellman operator. With $x=v_c$, the vector $v$ satisfies the Bellman equations of $G_c(x)$, including the new self-loop equation at $c$. Uniqueness gives $U(v_c)=v$, and the original equation at $c$ gives $g(v_c)=0$. Strict decrease proves uniqueness of the root and the remaining assertions.
\end{proof}

\subsubsection{Eliminating a random vertex}\label{sec:recursive-procedure}

Algorithm~\ref{alg:discounted} simulates the lower-level solver at $X=x^*$, carrying each reward register as $aX+b$. Coefficient and discrete instructions run directly. We resolve each comparison by the rule below, maintaining agreement with the run at $x^*$.

\begin{algorithm}[H]
\caption{Discounted policies by random-vertex elimination}\label{alg:discounted}
\small
\begin{algorithmic}[1]
\Require $G$, common discount $0<\beta<1$, and edge rewards $d$
\Ensure Uniformly optimal discounted policies $\sigma,\tau$
\If{$V_R=\varnothing$}
  \State \Return $\Call{DeterministicPolicies}{G,\beta,d}$
\EndIf
\State Choose $c\in V_R$ and form the terminal instance $G_c(X)$
\State Symbolically run $\Call{DiscountedPolicies}{G_c,\beta,d^{(X)}}$ using the rule below
\For{each comparison $\sgn(aX+b)$ requested by this run}
  \If{$a=0$}
    \State $s\gets\sgn(b)$
  \Else
    \State $y\gets-b/a$ and $(\sigma_y,\tau_y)\gets\Call{DiscountedPolicies}{G_c,\beta,d^{(y)}}$
    \State $w\gets\Call{EvaluatePair}{G_c,d^{(y)},\sigma_y,\tau_y,\beta}$
    \State $s\gets\sgn(a)\sgn\!\left(\sum_{i(e)=c}p_e(d_e+\beta w_{j(e)})-y\right)$
  \EndIf
  \State Supply $s$ and continue the symbolic run
\EndFor
\State \Return the returned policies restricted to $V_{\max},V_{\min}$
\end{algorithmic}
\end{algorithm}

Trials use one fewer random vertex and the linear evaluation of Lemma~\ref{lem:evaluation}. Their residual uses the original data at $c$; only its sign, including zero, returns to the simulation.

\begin{proposition}\label{prop:elimination}
Suppose $A_{k-1}$ returns uniformly optimal discounted policies for games with at most $k-1$ random vertices, has the affine register structure of Section~\ref{sec:affine}, and uses at most $T_{k-1}(N)$ operations uniformly in the numerical input. Algorithm~\ref{alg:discounted} extends it to $k$ random vertices with the same structure and
$$
T_k(N)\le N^b\bigl(T_{k-1}(N)+1\bigr)^2
$$
for an absolute constant $b$. Uniform polynomial bounds on affine coefficient lengths are preserved for each fixed $k$.
\end{proposition}
\begin{proof}
The simulation follows the run at $x^*$. A comparison has the form $a x+b\gtreqless0$, with reward-independent slope $a$ and reward-affine intercept $b$ on the current branch. For $a\ne0$, a trial at $y=-b/a$ evaluates $U(y)$, and Lemma~\ref{lem:root} gives $\sgn(a x^*+b)=\sgn(a)\sgn(g(y))$, including the zero case. For $a=0$, the sign is read directly. Induction over comparison requests therefore gives the policies optimal in $G_c(x^*)$. They are tight for $U(x^*)$ at every original controlled vertex. Restoring the chance equation at $c$ preserves this value vector by Lemma~\ref{lem:root}, so these policies are optimal in the original game.

There is at most one ordinary lower-level call and one linear policy evaluation per simulated comparison. Affine arithmetic in one formal parameter has polynomial overhead, and Lemma~\ref{lem:evaluation} gives polynomial cost for evaluation. This proves the recurrence. The affine structure is preserved because $y=-b/a$ is reward-affine, the lower-level program takes affine combinations of its substituted rewards, and policy evaluation uses a reward-independent matrix. In particular, its trial residual $g(y)$ is reward-affine on each branch.

For the bit bound, let $L_c$ be the complete length of the rational probabilities and discount, and let $C_{k-1}(N,L_c)$ bound the lower-level affine coefficients independently of its rewards. Substituting the terminal reward $(1-\beta)X$ bounds each comparison slope and each intercept coefficient by $\poly(N)(C_{k-1}+L_c+1)$ bits. Division by the reward-independent slope gives the same bound for the coefficients of $y$. Each register of an ordinary lower-level run is an affine form in at most $m$ rewards. Substitution therefore uses at most $m+1$ sums of products of bounded coefficients. The subsequent evaluation and residual have coefficients of polynomial length in $N,L_c$ by Lemma~\ref{lem:evaluation}. Thus universal constants $K,c$ give
$$
C_k(N,L_c)\le K N^c\bigl(C_{k-1}(N,L_c)+L_c+1\bigr).
$$
Only signs return from the trials, so numerical query outputs never accumulate through later queries. The symbolic execution itself follows one lower-level branch. The displayed bound is polynomial for fixed $k$; evaluating its affine forms at the original rational rewards also gives polynomial numerical bit lengths.
\end{proof}

\begin{corollary}\label{cor:discounted}
Algorithm~\ref{alg:discounted} returns uniformly optimal discounted policies for games with $k$ random vertices in $N^{O(2^k)}$ operations, uniformly over real rewards, transition probabilities, and common discounts $0<\beta<1$. The computation has the affine register structure and polynomial rational bit bounds for fixed $k$. Exact discounted values follow by one call to \textup{\textsc{EvaluatePair}} with the same asymptotic bounds.
\end{corollary}
\begin{proof}
Each eliminated random vertex is replaced by one deterministic self-loop, so all recursive arenas have size at most $N$. Starting with $T_0(N)\le N^a$ from Theorem~\ref{thm:base}, Proposition~\ref{prop:elimination} gives exponent $O(2^k)$ and the structural and bit assertions. Lemma~\ref{lem:evaluation} supplies the final optional evaluation.
\end{proof}

\subsubsection{A common stable discount}\label{sec:discount}

We now choose a discount preserving optimality near one. Andersson and Miltersen~\cite[Lemma~1]{AM} gave quantitative reductions using rational policy values and polynomial sign stabilization. Gaubert, Grand-Cl\'ement, and Katz~\cite{GGK} developed bounds for Blackwell and sensitive-optimality thresholds. We express all residual coefficients in one list of probability monomials and apply Lemma~\ref{lem:separation} to select the discount.

Let $E_i$ be the outgoing set of random vertex $c_i$ and put $d_i=|E_i|$. Adjoin a formal symbol $0$, set $p_0=1$, and define
$$
P_\alpha=\prod_{i=1}^k p_{\alpha_i},\qquad
z=\bigl(1,(r_eP_\alpha)_{e\in E,\ \alpha\in\prod_i(\{0\}\cup E_i)}\bigr).
$$
Repeated entries retain separate indices. For $k=0$ the product has one empty choice. The length satisfies $s=1+m\prod_i(1+d_i)\le1+m(m+1)^k$, so $z$ is computable in $N^{O(k+1)}$ operations. Lemma~\ref{lem:separation} implements \textsc{Separate}, which returns a lower bound on every nonzero integer combination of entries of $z$ with coefficient $1$-norm at most $B$.

\begin{algorithm}[H]
\caption{A discount preserving all policy inequalities}\label{alg:stable}
\small
\begin{algorithmic}[1]
\Require A BWR arena $G$ and rewards $r$
\Ensure A scalar $t_0\in(0,1/4]$ preserving policy optimality on $(0,t_0]$
\State Form $z$ as above and set $B\gets2^{16N^3}$
\State $\delta\gets\Call{Separate}{z,B}$
\State \Return $\min\{1/4,\delta/(4\max\{1,B\norm{z}_\infty\})\}$
\end{algorithmic}
\end{algorithm}

\paragraph{Policy residuals.}

Use the normalized discounted objective with $\beta=1-t$ and edge rewards $t r_e$, for $0<t<1$. Fix a policy pair $\pi$, let $P^\pi$ be its transition matrix, and let $r^\pi$ be its expected one-step reward vector. The policy value is
$$
v^\pi(t)=\bigl(I-(1-t)P^\pi\bigr)^{-1}t r^\pi.
$$
Write $D^\pi(t)=\det(I-(1-t)P^\pi)$ and let $C_j^\pi(t)$ be the Cramer numerator for coordinate $j$. The determinant $D^\pi(t)$ is positive for $0<t<1$. Indeed, $I-\beta P^\pi$ is nonsingular for $0\leq\beta<1$, and its determinant is continuous and equals one at $\beta=0$.

For a controlled edge $e$ with tail $i$, define the Bellman residual polynomial
$$
F_{\pi,e}(t)=t r_eD^\pi(t)+(1-t)C_{j(e)}^\pi(t)-C_i^\pi(t).
$$
The pair $\pi$ is discounted optimal exactly when all these residuals are nonpositive at Max vertices and nonnegative at Min vertices. Chance equations and the chosen-edge equations already hold by policy evaluation.

\begin{lemma}\label{lem:coefficients}
Every coefficient of every polynomial $F_{\pi,e}$ is an integer linear combination of the entries of $z$. A uniform bound on the $1$-norm of the integer coefficient vector is
$$
B=2^{16N^3}.
$$
Each residual polynomial has degree at most $n+1$.
\end{lemma}

\begin{proof}
Expand a determinant by permutations and then expand its entries. Every determinant term takes one entry from each row. At a controlled row, the transition is deterministic and contributes no probability variable. At a random row $c_i$, a chosen entry contributes either no probability or one outgoing probability $p_e$. If that row supplies an entry from the reward column in a Cramer determinant, the contribution is $t p_e r_e$ for one outgoing edge. Thus at most one outgoing probability is selected from each random row.

Each term of a Cramer numerator contains exactly one reward. Multiplication of $D^\pi$ by $r_e$ for a controlled edge also introduces exactly one reward and uses no additional random row. Consequently every residual coefficient is an integer combination of the listed entries $r_eP_\alpha$.

Each matrix entry has degree at most one in $t$. Before coefficients are collected, it has at most $1+2m$ monomial terms, including the diagonal contribution. The same upper bound covers reward-column entries. There are $n!$ permutation terms in a determinant. The total absolute coefficient mass of a determinant expansion is therefore at most $n!(1+2m)^n$. The two additional multiplications in a residual increase the bound by at most a factor four. The larger bound $4n!(2m+2)^{n+1}$ covers all residuals. For $N=2+n+m$, it is smaller than $2^{16N^3}$. Collecting terms can only decrease this absolute coefficient-mass bound. The degree bound follows from the same expansion.
\end{proof}

\paragraph{Blackwell optimality.}

Algorithm~\ref{alg:stable} uses this common coefficient bound to select a discount. Write $U=\max\{1,B\norm{z}_\infty\}$ for the coefficient magnitude bound.

\begin{proposition}\label{prop:blackwell}
Let $t_0$ be returned by Algorithm~\ref{alg:stable}. Every policy pair optimal for the normalized discounted game at $t=t_0$ is optimal for every $0<t\leq t_0$. Such a pair is uniformly optimal for the original mean-payoff game. Computing $t_0$ takes $N^{O(k+1)}$ operations. For each fixed $k$, its rational intermediate numbers have polynomial bit length in the complete input.
\end{proposition}

\begin{proof}
Every residual coefficient has magnitude at most $U$, and every nonzero coefficient has magnitude at least $\delta$. Let $F(t)=\sum_{h=\ell}^{D}a_ht^h$ be a nonzero residual polynomial, with $a_\ell\ne0$. For $0<t\leq t_0$,
$$
\abs{\sum_{h>\ell}a_ht^{h-\ell}}
\leq\frac{Ut}{1-t}\leq\frac{\delta}{3}<|a_\ell|.
$$
Hence $\sgn F(t)=\sgn a_\ell$ throughout this interval. An identically zero residual stays zero. Every policy pair therefore has the same Bellman inequality signs throughout the interval. A pair optimal at $t_0$ remains optimal throughout it.

The resulting pair is Blackwell optimal for the family with common discount $1-t$. The classical perfect-information Blackwell transfer implies that it is uniformly optimal for mean payoff~\cite{GZ}. The transfer applies to the expected long-run objective and to arbitrary opposing strategies. To match a state-reward formulation, subdivide every edge by an intermediate state of reward $2r_e$, assign zero reward to original states, and use a common discount $\gamma$ satisfying $\gamma^2=1-t$. The new normalized discounted objective is the old one multiplied by $2\gamma/(1+\gamma)>0$, and the mean payoff is unchanged. The liminf formulation follows by applying the transfer to the negated rewards and exchanged player roles, together with the limsup formulation.

The vector dimension satisfies $s\leq1+m(m+1)^k$, and $\log B=16N^3$. Constructing its entries and running Lemma~\ref{lem:separation} cost $N^{O(k+1)}$ operations. Products forming $z$ have at most $k+1$ scalar factors. The separation lemma, the computation of $U$, and the final division defining $t_0$ all have polynomial rational bit bounds for each fixed $k$.
\end{proof}

\subsection{Description of the main algorithm}\label{sec:algorithm}\label{sec:main-proof}

The main algorithm combines discount selection, the recursive discounted solver, and policy evaluation. Positive reward scaling preserves optimal policies, so the discounted search uses the original rewards $r$.

\begin{algorithm}[H]
\caption{Exact mean-payoff values and policies}\label{alg:mean}
\small
\begin{algorithmic}[1]
\Require A BWR arena $G$ and real edge rewards $r$
\Ensure The mean-payoff value vector $v$ and uniformly optimal policies $\sigma,\tau$
\State $t_0\gets\Call{StableDiscount}{G,r}$
\State $(\sigma,\tau)\gets\Call{DiscountedPolicies}{G,1-t_0,r}$
\State $v\gets\Call{EvaluatePair}{G,r,\sigma,\tau}$
\State \Return $(v,\sigma,\tau)$
\end{algorithmic}
\end{algorithm}

\begin{proof}[Proof of Theorem~\ref{thm:main}]
Algorithm~\ref{alg:mean} calls \textsc{StableDiscount}, whose positive output $t_0$ satisfies Proposition~\ref{prop:blackwell}. Corollary~\ref{cor:discounted} proves that \textsc{DiscountedPolicies} returns policies optimal at discount $1-t_0$ with rewards $r$. Positive scaling by $t_0$ preserves their optimality, so Proposition~\ref{prop:blackwell} makes them uniformly optimal for mean payoff. Finally, Lemma~\ref{lem:evaluation} proves that \textsc{EvaluatePair} returns their mean-payoff vector. Uniform optimality identifies that vector with the game value from every initial vertex.

Discount selection costs $N^{O(k+1)}$ operations, the discounted solution costs $N^{O(2^k)}$, and policy evaluation costs a polynomial in $N$. Since $k+1=O(2^k)$, their sum is $N^{O(2^k)}$. For fixed $k$, Proposition~\ref{prop:blackwell} bounds the rational size of the selected discount by a polynomial in the original input length. The bit assertion in Corollary~\ref{cor:discounted} and the bound in Lemma~\ref{lem:evaluation} then establish the bit assertion of the theorem.
\end{proof}


\begin{thebibliography}{99}

\bibitem{AGKS}
X. Allamigeon, S. Gaubert, R. D. Katz, and M. Skomra.
\newblock Universal complexity bounds based on value iteration for stochastic mean payoff games and entropy games.
\newblock \emph{Information and Computation}, 302, article 105236, 2025.
\newblock \url{https://doi.org/10.1016/j.ic.2024.105236}.

\bibitem{AM}
D. Andersson and P. B. Miltersen.
\newblock The complexity of solving stochastic games on graphs.
\newblock In \emph{ISAAC 2009}, Lecture Notes in Computer Science, 5878, 112--121, 2009.
\newblock \url{https://doi.org/10.1007/978-3-642-10631-6_13}.

\bibitem{ACS}
D. Auger, P. Coucheney, and Y. Strozecki.
\newblock Solving simple stochastic games with few random nodes faster using Bland's rule.
\newblock In \emph{STACS 2019}, LIPIcs, 126, 9:1--9:16, 2019.
\newblock \url{https://doi.org/10.4230/LIPIcs.STACS.2019.9}.

\bibitem{BEGMcanonical}
E. Boros, K. Elbassioni, V. Gurvich, and K. Makino.
\newblock On canonical forms for zero-sum stochastic mean payoff games.
\newblock \emph{Dynamic Games and Applications}, 3(2), 128--161, 2013.
\newblock \url{https://doi.org/10.1007/s13235-013-0075-x}.

\bibitem{BEGMdiscount}
E. Boros, K. Elbassioni, V. Gurvich, and K. Makino.
\newblock On discounted approximations of undiscounted stochastic games and Markov decision processes with limited randomness.
\newblock \emph{Operations Research Letters}, 41(4), 357--362, 2013.
\newblock \url{https://doi.org/10.1016/j.orl.2013.04.006}.

\bibitem{BEGMconvex}
E. Boros, K. Elbassioni, V. Gurvich, and K. Makino.
\newblock A convex programming-based algorithm for mean payoff stochastic games with perfect information.
\newblock \emph{Optimization Letters}, 11(8), 1499--1512, 2017.
\newblock \url{https://doi.org/10.1007/s11590-017-1140-y}.

\bibitem{BEGM}
E. Boros, K. Elbassioni, V. Gurvich, and K. Makino.
\newblock A pseudo-polynomial algorithm for mean payoff stochastic games with perfect information and few random positions.
\newblock \emph{Information and Computation}, 267, 74--95, 2019.
\newblock \url{https://doi.org/10.1016/j.ic.2019.03.005}.
\newblock Earlier version: \href{https://arxiv.org/abs/1508.03431v2}{arXiv:1508.03431v2}, 2017.

\bibitem{BEFGMM}
E. Boros, K. Elbassioni, M. Fouz, V. Gurvich, K. Makino, and B. Manthey.
\newblock Approximation schemes for stochastic mean payoff games with perfect information and few random positions.
\newblock \emph{Algorithmica}, 80, 3132--3157, 2018.
\newblock \url{https://doi.org/10.1007/s00453-017-0372-7}.

\bibitem{Condon}
A. Condon.
\newblock The complexity of stochastic games.
\newblock \emph{Information and Computation}, 96(2), 203--224, 1992.
\newblock \url{https://doi.org/10.1016/0890-5401(92)90048-K}.

\bibitem{FT}
A. Frank and \texorpdfstring{\'E.}{E.} Tardos.
\newblock An application of simultaneous Diophantine approximation in combinatorial optimization.
\newblock \emph{Combinatorica}, 7(1), 49--65, 1987.
\newblock \url{https://doi.org/10.1007/BF02579200}.

\bibitem{GGK}
S. Gaubert, J. Grand-Cl\'ement, and R. D. Katz.
\newblock Thresholds for sensitive optimality and Blackwell optimality in stochastic games.
\newblock \emph{arXiv:2506.18545}, 2025.
\newblock \url{https://arxiv.org/abs/2506.18545}.

\bibitem{GH}
H. Gimbert and F. Horn.
\newblock Solving simple stochastic games with few random vertices.
\newblock \emph{Logical Methods in Computer Science}, 5(2:9), 1--17, 2009.
\newblock \url{https://doi.org/10.2168/LMCS-5(2:9)2009}.

\bibitem{GZ}
H. Gimbert and W. Zielonka.
\newblock Blackwell-optimal strategies in priority mean-payoff games.
\newblock In \emph{Proceedings of GandALF 2010}, Electronic Proceedings in Theoretical Computer Science, 25, 7--21, 2010.
\newblock \url{https://doi.org/10.4204/EPTCS.25.5}.

\bibitem{GLS}
M. Gr\"otschel, L. Lov\'asz, and A. Schrijver.
\newblock \emph{Geometric Algorithms and Combinatorial Optimization}.
\newblock Algorithms and Combinatorics, vol.~2, Springer, second corrected edition, 1993.
\newblock \url{https://doi.org/10.1007/978-3-642-78240-4}.

\bibitem{GKK}
V. A. Gurvich, A. V. Karzanov, and L. G. Khachiyan.
\newblock Cyclic games and an algorithm to find minimax cycle means in directed graphs.
\newblock \emph{USSR Computational Mathematics and Mathematical Physics}, 28(5), 85--91, 1988.
\newblock \url{https://doi.org/10.1016/0041-5553(88)90012-2}.

\bibitem{HMZ}
T. D. Hansen, P. B. Miltersen, and U. Zwick.
\newblock Strategy iteration is strongly polynomial for 2-player turn-based stochastic games with a constant discount factor.
\newblock \emph{Journal of the ACM}, 60(1), 1:1--1:16, 2013.
\newblock \url{https://doi.org/10.1145/2432622.2432623}.

\bibitem{IJM}
R. Ibsen-Jensen and P. B. Miltersen.
\newblock Solving simple stochastic games with few coin toss positions.
\newblock In \emph{ESA 2012}, Lecture Notes in Computer Science, 7501, 636--647, 2012.
\newblock \url{https://doi.org/10.1007/978-3-642-33090-2_55}.

\bibitem{LLL}
A. K. Lenstra, H. W. Lenstra, Jr., and L. Lov\'asz.
\newblock Factoring polynomials with rational coefficients.
\newblock \emph{Mathematische Annalen}, 261, 515--534, 1982.
\newblock \url{https://doi.org/10.1007/BF01457454}.

\bibitem{LL}
T. M. Liggett and S. A. Lippman.
\newblock Stochastic games with perfect information and time average payoff.
\newblock \emph{SIAM Review}, 11(4), 604--607, 1969.
\newblock \url{https://doi.org/10.1137/1011093}.

\bibitem{OpenAIgames}
OpenAI.
\newblock Turn-based stochastic mean-payoff games in deterministic quasipolynomial time.
\newblock Preprint, October 5, 2026.
\newblock \href{https://github.com/openai/math/blob/main/preprints/Turn-Based-Stochastic-Mean-Payoff-Games-in-Deterministic-Quasipolynomial-Time-October-5-2026/stochastic-mean-payoff-games.pdf}{Manuscript}.

\bibitem{OpenAI}
OpenAI.
\newblock Strongly polynomial linear programming.
\newblock Preprint, November 24, 2026.
\newblock \href{https://github.com/openai/math/tree/main/preprints/Strongly-polynomial-linear-programming-November-24-2026}{Manuscript and source}.

\bibitem{PY}
I. Post and Y. Ye.
\newblock The simplex method is strongly polynomial for deterministic Markov decision processes.
\newblock \emph{Mathematics of Operations Research}, 40(4), 859--868, 2015.
\newblock \url{https://doi.org/10.1287/moor.2014.0699}.

\bibitem{ZP}
U. Zwick and M. Paterson.
\newblock The complexity of mean payoff games on graphs.
\newblock \emph{Theoretical Computer Science}, 158(1--2), 343--359, 1996.
\newblock \url{https://doi.org/10.1016/0304-3975(95)00188-3}.

\end{thebibliography}
\end{document}